\documentstyle[amsfonts,12pt]{article}

\def\half{\textstyle{\frac{1}{2}}}

\def\H{{\cal H}}
\def\D{{\cal D}}

\def\E{{\rm I}\hskip-.2em{\rm E}}
\def\ra{\rightarrow}
\def\tint{{\textstyle\int}}
\def\hg{{\hat g}}
\def\hp{{\hat\pi}}

\def\s{\hskip.08em}
\def\l{\lambda}

\def\o{\overline}
\def\a{\alpha}
\def\b{\begin{eqnarray*}}     
\def\e{\end{eqnarray*}}       
\def\bn{\begin{eqnarray}}     
\def\en{\end{eqnarray}}       
\def\<{\langle}
\def\>{\rangle}

\def\no{\nonumber}

\def\dnx{d^{n}\!x}

\def\{{\lbrace}
\def\}{\rbrace}
\begin{document}
\title{A Coherent Strategy for Quantum Gravity\footnote{Presented at the 
XXV International Colloquium on Group Theoretical Methods 
in Physics, Cocoyoc, Mexico, August, 2004}}
\author{John R. Klauder
\footnote{Electronic mail: klauder@phys.ufl.edu}\\
Departments of Physics and Mathematics\\
University of Florida\\
Gainesville, FL  32611}
\date{}     
\maketitle
\begin{abstract}
Affine quantum gravity, which differs notably from either string theory or
loop quantum gravity, is briefly reviewed. Emphasis in this article is
placed on the use of affine coherent states in this program.
\end{abstract}
\section*{Introduction}
Quantum gravity is more difficult than other quantum field theories because
of (i) metric positivity,  (ii) partially second-class quantum constraints,
and (iii) perturbative nonrenormalizability, to mention just a few
problems. In this paper we briefly review several of these issues.
\subsection*{Metric positivity} 
Distinct points in a space-like 3-dimensional manifold have a positive 
separation distance. As a consequence the spatial metric $g_{ab}(x)$ 
satisfies the requirement that $u^a\s g_{ab}(x)\s u^b>0$ whenever $u^a$ 
is not identically zero.
 We insist that the associated quantum operator $\hg_{ab}(x)$ 
also satisfy metric positivity such that $u^a\s\hg_{ab}(x)\s u^b>0$ in
the sense of operators for all nonvanishing $u^a$. Moreover, we insist 
that $\hg_{ab}(x)$ becomes self adjoint when smeared with a suitable
real test function. In canonical quantization one chooses the canonical 
(ADM) momentum $\pi^{ab}(x)$ as the field to promote to an operator, 
$\hp^{ab}(x)$. 
However, since the momentum acts to {\it translate} the metric, 
such a choice is
inconsistent with the preservation of metric positivity. Instead, it is
appropriate to choose the {\it mixed-valence momentum field} $\pi^a_c(x)\equiv 
\pi^{ab}(x)\s g_{bc}(x)$ to promote to an operator, $\hp^a_c(x)$. The operator
$\hp^a_c(x)$, along with the operator for the metric $\hg_{ab}(x)$, fulfill
the {\it affine commutation relations} given (for $\hbar=1$) by 
\cite{klagra} 
 \b 
  &&\hskip.2cm[\hp^a_b(x),\,\hp^c_d(y)]=\half\s i\s[\s\delta^c_b\hp^a_d(x)
-\delta^a_d\hp^c_b(x)\s]\,\delta(x,y)\;,\no\\
  &&\hskip.1cm[\hg_{ab}(x),\,\hp^c_d(y)]=\half\s i\s[\s\delta^c_a\hg_{bd}(x)
+\delta^c_b\hg_{ad}(x)\s]\,\delta(x,y)\;,\label{afc}\\
&&[\hg_{ab}(x),\,\hg_{cd}(y)]=0\;.   \no \e
When smeared with real test functions, both $\hp^a_c(x)$ and $\hg_{ab}(x)$ 
become self-adjoint operators.
The appropriateness of the affine commutation relations
is confirmed by the relation
  \b e^{i\tint \gamma^a_b(y)\s\hp^b_a(y)\s d^3\!y}\,\hg_{cd}(x)\,
e^{-i\tint \gamma^a_b(y)\s\hp^b_a(y)\s d^3\!y}=(\s e^{\gamma(x)/2}\s)^e_c\;
\hg_{ef}(x)\,(\s e^{\gamma(x)/2}\s)^f_d\;,  \e
which clearly preserves metric positivity.

A representation of the basic kinematical operator fields is encoded
in a choice of {\it affine coherent states} defined by
\b
  |\pi,g\>\equiv e^{i\tint \pi^{ab}(x)\,\hg_{ab}(x)\,d^3\!x}\,
e^{-i\tint\gamma^d_c(x)\,{\hat\pi}^c_d(x)\,d^3\!x}\,|\eta\>\;,  \e
for all smooth fields $\pi$ and $\gamma$; the reason for the choice of 
the ket label will become clear below. As 
minimum requirements on the fiducial vector $|\eta\>$ we impose
 $\hskip.08cm\<\eta|\s\hp^c_d(x)\s|\eta\>\equiv 0$ and 
$\<\eta|\s\hg_{ab}(x)\s|\eta\>\equiv{\tilde g}_{ab}(x)$.  
Here, ${\tilde g}(x)\equiv\{{\tilde g}_{ab}(x)\}$ is a fixed, smooth, 
positive-definite metric function determined by the choice of $|\eta\>$. 
The form of ${\tilde g}(x)$ determines the topology of the space-like 
surface under consideration; if that surface is noncompact, then 
${\tilde g}(x)$ also determines its asymptotic nature.

It follows that
  \b  &&\<\pi,g|\s\hg_{cd}(x)\s|\pi,g\>\equiv g_{cd}(x)
\equiv(\s e^{\gamma(x)/2}\s)^e_c\;
{\tilde g}_{ef}(x)\,(\s e^{\gamma(x)/2}\s)^f_d\;,  \\
   &&\hskip.09cm\<\pi,g|\s\hp^a_c(x)\s|\pi,g\>\equiv\pi^a_c(x)
\equiv\pi^{ab}(x)\s g_{bc}(x)\;. \e

The full specification of $|\eta\>$, and thereby of the representation of
the affine field operators, is implicitly given by the form of the overlap 
function of
two coherent states
 \b &&\hskip-.4cm\<\pi'',g''|\pi',g'\>=\exp\bigg[-2\int b(x)\,d^3\!x\,\no\\
&&\hskip.1cm\times\ln\bigg(\frac{\det\{\half[g''^{ab}(x)+g'^{ab}(x)]+
\half ib(x)^{-1}[\pi''^{ab}(x)-\pi'^{ab}(x)]\}}{\{\det[g''^{ab}(x)]\,
\det[g'^{ab}(x)]\}^{1/2}}\bigg)\bigg]\;. \e
As is evident, this expression depends only on the six components of $g_{ab}$ 
rather than the nine components of $\gamma^a_b$; thus it is adequate
to label the coherent states by $\pi, g$ rather than $\pi,\gamma$.

As a further general comment about  $\<\pi'',g''|\pi',g'\>$ we observe 
that it is {\it invariant} under general (smooth, invertible) coordinate 
transformations $x\ra {\o x}={\o x}(x)$, and we say that the given 
expression characterizes a {\it diffeomorphism covariant realization} of 
the affine field operators. This property holds, in part, because $b(x)$, 
restricted so that $0<b(x)<\infty$,  transforms as a {\it scalar density} 
in both places that it appears. 
Thus $b(x)$, which has the dimensions of ${\sf L}^{-3}$, plays an essential 
dimensional and transformational role. 

It is particularly significant that the coherent state overlap functional 
also admits a functional integral formulation given by
\b  &&\<\pi'',g''|\pi',g'\>
=\lim_{\nu\ra\infty}\,{\o{\cal N}}_{\nu}\,\int 
\exp[-i\tint g_{ab}\s{\dot\pi}^{ab}\,d^3\!x\,dt]\no\\
  &&\hskip.6cm\times\exp\{-(1/2\nu)\tint[b(x)^{-1}
g_{ab}g_{cd}{\dot\pi}^{bc}{\dot\pi}^{da}+b(x)g^{ab}g^{cd}{\dot g}_{bc}
{\dot g}_{da}]\,d^3\!x\,dt\}\no\\
&&\hskip2.6cm\times\Pi_{x,t}\,\Pi_{a\le b}\,d\pi^{ab}(x,t)\,dg_{ab}(x,t) 
\label{e20}\;.  \e
Observe that this expression involves a continuous-time regularization factor
that already gives it an essentially rigorous definition. A continuous-time 
regularization is an alternative to a lattice regularization commonly used 
to define such expressions. For further details see \cite{klagra}.
\subsection*{Gravitational constraints}
Let us proceed formally in order to see the essence of the quantum constraint 
problem. Suppose that $\H_a(x)$ and $\H(x)$ represent local self-adjoint 
constraint operators for the gravitational field. Standard calculations 
lead to the commutation relations 
 \b  && [\H_a(x),\H_b(y)]=i\s[\delta_{,a}(x,y)\,\H_b(x)-\delta_{,b}(x,y)\,
\H_a(x)]\;, \\
    && \hskip.18cm[\H_a(x),\H(y)]=i\s\delta_{,a}(x,y)\,\H(x)\;,  \\
 && \hskip.36cm[\H(x),\H(y)]= i\s\half\s\delta_{,a}(x,y)[\,\hg^{ab}(x)\,
\H_b(x)+\H_b(x)\,\hg^{ab}(x)\,]\;.  \e
Based on the last of these expressions one finds that the quantum 
gravitational constraints are partially second class. As such their 
treatment causes additional complications. 

The recently developed {\it projection operator method} of constraint 
quantization \cite{proj} is well suited to the quantization of such systems. 
Briefly stated, if $\{\Phi_\a\}_{\a=1}^A$ denotes a set of self-adjoint
quantum constraint operators, it is satisfactory to consider the
regularized physical Hilbert space ${\frak H}_{phys}\equiv\E\s{\frak H}$, 
where $\frak H$ denotes the original Hilbert space and $\E$ is a
projection operator which may be defined by
 \b \E(\Sigma_\a\Phi^2_\a\le\delta(\hbar)^2)=\int {\bf T}\,
e^{-i\tint\lambda^\a(t)\s\Phi_\a\,dt}\,\D R(\lambda)\;. \e
Here ${\bf T}$ denotes the time-ordering operator, 
$\{\lambda^\a(t)\}_{\a=1}^A$, $0\le t< T$, denotes a set of 
$c$-number ``Lagrange multiplier'' functions, and $\D R(\lambda)$ 
denotes a formal measure on such functions, which
is decidedly not a ``flat'' measure. Observe that $\E$ projects onto a 
small interval of the spectral domain of $\Sigma_\a\Phi^2_\a$ bounded by 
zero and $\delta(\hbar)^2$. Suitable choices of $\delta(\hbar)^2$, and 
possible limits when
$\delta\ra0$, are discussed in \cite{proj}.

Applied to the gravitational problem, such ideas lead to 
\b  && \<\pi'',g''|\s\E\s|\pi',g'\>  
   =\lim_{\nu\ra\infty}{\o{\cal N}}_\nu\s\int 
e^{-i\tint[g_{ab}{\dot\pi}^{ab}+N^aH_a+NH]\,d^3\!x\,dt}\no\\
  &&\hskip1.5cm\times\exp\{-(1/2\nu)\tint[b(x)^{-1}
g_{ab}g_{cd}{\dot\pi}^{bc}{\dot\pi}^{da}+b(x)g^{ab}g^{cd}{\dot g}_{bc}
{\dot g}_{da}]\,d^3\!x\,dt\}\no\\
  &&\hskip2cm\times\bigg[\Pi_{x,t}\,\Pi_{a\le b}\,d\pi^{ab}(x,t)\,
dg_{ab}(x,t)\bigg]\,\D R(N^a,N)\;. \e
This important formal expression realizes the coherent-state matrix 
elements of the desired projection operator, which then represents a 
quantity that may be used as a {\it reproducing kernel} to define the 
physical Hilbert
space as a reproducing kernel Hilbert space; see \cite{klagra}. 

\subsection*{Perturbative nonrenormalizability}
In order to resolve the problem of perturbative nonrenormalizability, it is
important to understand the cause of such behavior. We claim this behavior
can be understood from the 
{\it hard-core picture} of nonrenormalizable interactions which 
we now outline \cite{book}.

It is pedagogically useful to first examine singular potentials in quantum 
mechanics. In particular, 
consider the Euclidean-space path integral for a free particle in the
presence of a singular potential given by
 \b W_\l\equiv {\cal N}\int_{x(0)=x'}^{x(T)=x''}\, e^{-\half\tint 
{\dot x}(t)^2\s dt-\l\tint x(t)^{-4}\s dt}\;{\cal D}x \;. \label{e37}\e
As $\l\ra0^+$, it appears self evident that $W_\l$  passes to the
expression
 \b W_0\equiv{\cal N}\int_{x(0)=x'}^{x(T)=x''}\, e^{-\half\tint{\dot x}^2
(t)\s dt}\;{\cal D}x=\frac{1}{\sqrt{2\pi T}}\,e^{-(x''-x')^2/2T}  \e
appropriate to a free particle. Whatever the  dependence of 
$W_\l-W_0$ on small $\l$, it is tacitly
assumed that as $\l\ra0^+$, $W_\l\ra W_0$, i.e., that $W_\l$ is {\it 
continuously connected} to $W_0$. In this case, however, this expectation 
is incorrect.

In particular, when $\lambda>0$, the singularity at $x=0$ 
is so strong that the contribution from all paths that reach or cross the 
origin is {\it completely suppressed} since $\tint x(t)^{-4}\s dt=\infty$ 
for such
paths. As a consequence, as $\l\ra0^+$, 
it follows that
\b \lim_{\l\ra0^+}W_\l=W'_0\equiv\frac{\theta(x''x')}{\sqrt{2\pi T}}
\bigg[\s e^{-(x''-x')^2/2T}-e^{-(x''+x')^2/2T}\s\bigg]\;.  \e
Stated otherwise, $W_\l$ is {\it decidedly not} 
continuously connected to the
free theory $W_0$, but is instead continuously connected to an alternative
theory -- called a pseudofree theory -- that accounts for 
the {\it hard-core} 
effects of the interaction. The interacting theory may well possess a 
perturbation expansion about the pseudofree theory, but the interacting 
theory will {\it not} possess any 
perturbation expansion about the free theory. 

Next consider a scalar field theory and the Euclidean-space functional 
integral 
 \b S_\l(h)\equiv {\cal N}\int \exp\{{\tint} {h\phi\s\dnx}-\half
\tint[(\nabla\phi)^2+m^2\phi^2]\s\dnx-\l\tint\phi^4\s\dnx\}\;{\cal D}\phi  \e
appropriate to the $\phi^4_n$ model in $n$  spacetime dimensions.  We recall
for such expressions that there is a Sobolev-type inequality to the effect
that
  \b  \{\tint\phi(x)^4\s\dnx\}^{1/2}\le K\tint[(\nabla\phi(x))^2+m^2\phi(x)^2]
\s\dnx \e
holds for {\it finite} $K$ (e.g., $K=4/3$) whenever $n\le 4$, but which 
{\it fails} to hold (i.e., $K=\infty$) whenever $n\ge5$. Thus for 
nonrenormalizable
interactions $\phi^4_n$, for which $n\ge5$, it follows that there are fields
$\phi$ for which the free action is finite while the interaction action is 
infinite. This is the formal criterion for hard-core behavior.

Lastly we observe that gravity is also a theory for which the free action 
(limited to quadratic terms) does not dominate the interaction action 
(remaining terms),
and consequently gravity would seem to be a candidate theory to be 
understood on the basis of a hard-core interaction. Work in this
direction proceeds.

\end{document}